\documentstyle[a4,12pt]{article}

\newcommand{\be}{\begin{equation}}
\newcommand{\ee}{\end{equation}}
\newcommand{\ba}{\begin{eqnarray}}
\newcommand{\ea}{\end{eqnarray}}

\begin{document}
\pagestyle{plain}
\begin{center}
{\bf A COLLECTIVE MODEL OF BARYONS}\\
\end{center}
\mbox{}
\begin{center}
R.~Bijker$^{(1)}$, F. Iachello$^{(2)}$ and A.~Leviatan$^{(3)}$\\
\end{center}
\mbox{}\\
\indent \hspace{1.9truecm} $^1$ R.J.~Van de Graaff Laboratory,
University of Utrecht,\\
\indent \hspace{2.2truecm} P.O. Box 80000, 3508 TA Utrecht, The Netherlands\\
\indent \hspace{1.9truecm} $^2$ Center for Theoretical Physics,
Sloane Laboratory,\\
\indent \hspace{2.2truecm} Yale University, New Haven, Connecticut 06511\\
\indent \hspace{1.9truecm} $^3$ Racah Institute of Physics,
The Hebrew University,\\
\indent \hspace{2.2truecm} Jerusalem 91904, Israel\\
\mbox{}\\
\mbox{}
\begin{center}
{\bf ABSTRACT}\\
\end{center}
We propose an algebraic description of the geometric structure of baryons in
terms of the algebra $U(7)$. We construct a mass operator that preserves the
threefold permutational symmetry and discuss a collective model of baryons
with the geometry of an oblate top.\\
\mbox{}\\
\mbox{}\\
PACS numbers: 14.20-c, 14.20.Dh, 14.20.Gk, 03.65.Fd
\newpage

The ongoing construction of advanced accelerators in the energy range
of 1--10 GeV is generating renewed interest in the spectroscopy of baryons.
A detailed description of the structure of baryons requires a treatment of
both the internal (spin-flavor-color) and geometric degrees of freedom.
The purpose of this note is twofold: (i) to introduce a $U(7)$ algebra in
order to provide a unified framework in which the geometric structure of
baryons can be analyzed and (ii) to discuss, within this framework, a
collective model of baryons in which the constituent parts (quarks or
otherwise) move in a correlated fashion, in contrast with the
single-particle picture of the quark potential model, either in its
nonrelativistic \cite{IK} or relativized \cite{CI} form.
The algebra $U(7)$ enlarges the scope of the algebraic models of a
generation ago \cite{DG} since, within its framework, it is relatively simple
to incorporate the main aspect of baryon structure,
{\it i.e.} its triality aspect according to which three constituent parts
must be present in baryons. This leads to a threefold symmetry
which has to be imposed on all operators and wave functions.

In order to introduce the $U(7)$ algebra, we begin by noting that two vector
coordinates characterize the geometric structure of an object with three
constituent parts. These can be conveniently taken as
the Jacobi coordinates
\ba
\vec{\rho} &=& (\vec{r}_1 - \vec{r}_2)/\sqrt{2} ~,
\nonumber\\
\vec{\lambda} &=& (\vec{r}_1 + \vec{r}_2 -2\vec{r}_3)/\sqrt{6} ~,
\ea
where $\vec{r}_i$ denotes the coordinate of the $i$-th constituent
(Fig.~1). The algebraic model we propose is obtained by quantizing the Jacobi
coordinates in (1) and their associated momenta. To this end, we introduce
two vector bosons with angular momentum and parity,
$L^{\pi}=1^{-}$, and a scalar boson with $L^{\pi}=0^{+}$. We denote the
corresponding creation and annihilation operators by
$p^{\dagger}_{\rho,m}$, $p^{\dagger}_{\lambda,m}$, $p_{\rho,m}$,
$p_{\lambda,m}$ (vector bosons) and $s^{\dagger}$, $s$ (scalar boson). The
index $m$ represents the three components of the vector bosons, $m=0,\pm 1$.
The $U(7)$ algebra is generated by the $49$ bilinear products
$G_{\alpha,\alpha'}=b^{\dagger}_{\alpha}b_{\alpha^{\prime}}$, where
$b^{\dagger}_{\alpha}$ ($\alpha=1,\ldots ,7$) denotes the set of seven
creation operators and $b_{\alpha^{\prime}}$ the corresponding set of
annihilation operators.
This quantization procedure follows the general criterion of associating
an algebra $U(k+1)$ to a problem with $k$ degrees of
freedom (in the present case $k=6$) and assigning all states
to the totally symmetric representation $[N]$ of $U(k+1)$. Such a bosonic
quantization scheme has proved to be very useful in nuclear \cite{AI}
and molecular \cite{IAC} physics.

The next step is to construct the mass operator and other operators of
interest in terms of elements of $U(7)$.
The eigenstates of the mass operator must
have well defined transformation properties under the permutation of the
identical constituent parts. For nonstrange baryons, assuming
charge-independence, this implies that
the mass operator must be invariant under $S_{3}$. The
transformation properties of all algebraic operators under $S_3$
follow from those of $s^{\dagger}$, $p^{\dagger}_{\rho}$
and $p^{\dagger}_{\lambda}$.
There are three different symmetry classes for the permutation of
three objects: a symmetric one, $S$, an antisymmetric one, $A$, and a
two-dimensional class of mixed symmetry type, denoted usually by $M_{\rho}$
and $M_{\lambda}$. The latter has the same transformation properties of
the creation operators, $p^{\dagger}_{\rho}$ and
$p^{\dagger}_{\lambda}$. Alternatively, the three symmetry classes can
be labelled by the irreducible representations of the point group
$D_{3}$ (which is isomorphic to $S_3$) as $A_1$, $A_2$ and $E$,
respectively. The $s$-boson is a scalar under the permutation group.
It is now straightforward to find bilinear
combinations of bosonic operators that transform
irreducibly under the permutation group \cite{BL}.
We use the multiplication rules of $S_3$ to construct
all rotationally invariant one- and two-body terms in the $U(7)$ mass
operator that preserve parity and that are scalars under the permutation
group
\ba
\hat M^2_{U(7)} &=& \epsilon_s \, s^{\dagger}s - \epsilon_p \,
(p_{\rho}^{\dagger} \cdot \tilde{p}_{\rho}
+p_{\lambda}^{\dagger} \cdot \tilde{p}_{\lambda})
+ u_0 \, s^{\dagger} s^{\dagger} s s - u_1 \,
  s^{\dagger} ( p^{\dagger}_{\rho} \cdot \tilde{p}_{\rho}
+ p^{\dagger}_{\lambda} \cdot \tilde{p}_{\lambda} ) s
\nonumber\\
&& + v_0 \, \left[ ( p^{\dagger}_{\rho} \cdot p^{\dagger}_{\rho}
  + p^{\dagger}_{\lambda} \cdot p^{\dagger}_{\lambda} ) s s
  + s^{\dagger} s^{\dagger} ( \tilde{p}_{\rho} \cdot \tilde{p}_{\rho}
  + \tilde{p}_{\lambda} \cdot \tilde{p}_{\lambda} ) \right]
\nonumber\\
&& + \sum_{l=0,2} c_l \, \left[ ( p^{\dagger}_{\rho} p^{\dagger}_{\rho}
  - p^{\dagger}_{\lambda} p^{\dagger}_{\lambda} )^{(l)} \cdot
  ( \tilde{p}_{\rho} \tilde{p}_{\rho}
  - \tilde{p}_{\lambda} \tilde{p}_{\lambda} )^{(l)}
+ 4 \, ( p^{\dagger}_{\rho} p^{\dagger}_{\lambda})^{(l)} \cdot
       ( \tilde p_{\lambda} \tilde p_{\rho})^{(l)} \right]
\nonumber\\
&& + c_1 \, ( p^{\dagger}_{\rho} p^{\dagger}_{\lambda} )^{(1)} \cdot
( \tilde p_{\lambda} \tilde p_{\rho} )^{(1)}
+ \sum_{l=0,2} w_l \, ( p^{\dagger}_{\rho} p^{\dagger}_{\rho}
  + p^{\dagger}_{\lambda} p^{\dagger}_{\lambda} )^{(l)} \cdot
  ( \tilde{p}_{\rho} \tilde{p}_{\rho}
  + \tilde{p}_{\lambda} \tilde{p}_{\lambda} )^{(l)} ~,
\nonumber\\
\ea
with $\tilde{p}_{\lambda,m}=(-1)^{1-m} p_{\lambda,-m}$ and a similar
expression for $\tilde p_{\rho,m}$. The parentheses $(l)$ denote
angular momentum couplings.
Eq.~(2) represents the most general $S_3$-invariant
mass operator that one can construct up to quadratic terms in the
elements of $U(7)$. The spectrum of this mass operator can be obtained
by straightforward diagonalization in the basis provided by the
symmetric irreducible representation $[N]$ of $U(7)$. The value of $N$
determines the number of states in the model space and in view of confinement
is expected to be large. For a given $N$ the model space contains
the oscillator shells with $n=n_{\rho}+n_{\lambda}=0,1,\ldots,N$.

The mass operator of eq.~(2) spans a large class of possible
geometric models of baryons.
When $v_0=0$ the mass operator is diagonal in the harmonic oscillator
basis. The linear terms correspond to an oscillator frequency
of $\epsilon_p-\epsilon_s$ and the remaining terms represent
anharmonic contributions. This case is the algebraic analogue of the harmonic
oscillator quark model \cite{IK} and corresponds to the subalgebra
$U(6)\otimes U(1)$ of $U(7)$.
The $v_0$ term gives rise to a coupling between oscillator shells and
hence generates different models of baryon structure, which are of a
more collective nature ({\it i.e.} such that when the wave functions are
expanded in an oscillator basis, they are spread over many shells).
In the remaining part of this letter, we discuss one of these models,
corresponding to the geometric structure of an oblate symmetric top.

In order to understand the physical content of the otherwise abstract
interaction terms in eq.~(2), it is convenient to analyze them in more
intuitive geometric terms. Geometric shape variables can be associated with
algebraic models by introducing coherent (or intrinsic)
states \cite{ibm}. For bosonic systems, the ground state intrinsic wave
function can be written as a condensate
\ba
\mid N;c \rangle &=&
\frac{1}{\sqrt{N!}} \Bigl( b_c^{\dagger} \Bigr)^N \mid 0 \rangle ~.
\ea
The condensate bosons of $U(7)$ are
\ba
b_c^{\dagger} &=& (1+R^2)^{-1/2} \Bigl[ s^{\dagger}
+ r_{\rho} \, p_{\rho,0 }^{\dagger} + r_{\lambda} \sum_{m}
d^{(1)}_{m, 0}(\theta) \, p_{\lambda, m}^{\dagger} \Bigr] \, ~,
\ea
where $R^{2} = r_{\rho}^{2} + r_{\lambda}^{2}$. They
depend on three geometric variables, two lengths $r_{\rho},\,r_{\lambda}$
and the relative angle $\theta$, {\it i.e.}
$\vec{r}_{\rho}\cdot\vec{r}_{\lambda}=r_{\rho}r_{\lambda}\cos\theta$.
In general, the equilibrium (ground state) configuration of the condensate can
be obtained by calculating the expectation value of $\hat M^{2}_{U(7)}$ in
the condensate wave function and minimizing it with respect to
$r_{\rho},\,r_{\lambda}$ and $\theta$.
The $S_3$-invariant operator of eq.~(2)
supports a variety of equilibrium configurations: spherical,
linear and nonlinear. In this letter we study the rigid nonlinear shape
characterized by the equilibrium values $r_{\rho}=r_{\lambda}=
R/\sqrt{2} >0$ and $\theta = \pi/2$. These are precisely the conditions
satisfied by the Jacobi coordinates in eq.~(1) for an equilateral triangular
shape. The variables $\vec{r}_{\rho}$ and $\vec{r}_{\lambda}$ are obtained
from $\vec{\rho}$ and $\vec{\lambda}$ by dividing by a scale and are
therefore dimensionless.

In order to construct a mass formula corresponding to this situation,
we begin by decomposing $\hat M^{2}_{U(7)}$ into a vibrational and
a rotational part,
$\hat M^{2}_{U(7)}=\hat M^{2}_{\mbox{vib}} + \hat M^{2}_{\mbox{rot}}$.
The vibrational part is given by
\ba
\hat M^{2}_{\mbox{vib}} &=& \xi_1 \,
\Bigl ( R^2 \, s^{\dagger} s^{\dagger}
- p^{\dagger}_{\rho} \cdot p^{\dagger}_{\rho}
- p^{\dagger}_{\lambda} \cdot p^{\dagger}_{\lambda} \Bigr ) \,
\Bigl ( R^2 \, s s - \tilde{p}_{\rho} \cdot \tilde{p}_{\rho}
- \tilde{p}_{\lambda} \cdot \tilde{p}_{\lambda} \Bigr )
\nonumber\\
&& + \xi_2 \, \Bigl [
\Bigl ( p^{\dagger}_{\rho} \cdot p^{\dagger}_{\rho}
- p^{\dagger}_{\lambda} \cdot p^{\dagger}_{\lambda} \Bigr ) \,
\Bigl ( \tilde{p}_{\rho} \cdot \tilde{p}_{\rho}
- \tilde{p}_{\lambda} \cdot \tilde{p}_{\lambda} \Bigr )
+ 4 \, \Bigl ( p^{\dagger}_{\rho} \cdot p^{\dagger}_{\lambda} \Bigr ) \,
\Bigl ( \tilde p_{\lambda} \cdot \tilde p_{\rho} \Bigr ) \Bigr ] ~.\qquad
\ea
By construction it annihilates the equilibrium condensate and depends
on three parameters, $\xi_{1},\,\xi_{2}$ and $R^2$, which are related to
$u_{0},\,v_{0},\,c_0$ and $w_{0}$ in eq.~(2).

A standard analysis \cite{LK} of the vibrational excitations gives, in the
limit in which $N$ is large, a harmonic vibrational spectrum
\ba
M^{2}_{\mbox{vib}} &=& N \left[ \lambda_1 \, n_u
+ \lambda_2 \, (n_v + n_w) \right] ~,
\nonumber\\
\lambda_1 &=& 4 \, \xi_1 R^2 ~,
\nonumber\\
\lambda_2 &=& 4 \, \xi_2 R^2 (1+R^2)^{-1} ~.
\ea
The vibrations consist of a symmetric stretching ($u$),
an antisymmetric stretching ($v$) and a bending vibration ($w$).
The first two are radial excitations, whereas the third is an angular
mode which corresponds to oscillations in the angle $\theta$ between
the two Jacobi coordinates. The angular mode is degenerate with the
antisymmetric radial mode. These features show that
$\hat M^{2}_{\mbox{vib}}$ of eq.~(5)
describes the vibrational excitations of an oblate symmetric top.

In addition to vibrational excitations, there are rotational excitations.
The rotational excitations are labelled by the orbital
angular momentum, $L$, its projection on the threefold symmetry
axis, $K=0,1,\ldots,$ parity and the transformation properties
under the permutation group. For a given value of $K$, the states
have angular momentum $L=K,K+1,\ldots,$ and parity $\pi=(-)^K$.
For rotational states built on vibrations of type $A_1$
(symmetric under $S_3$) each $L$ state is single for $K=0$ and
twofold degenerate for $K \neq 0$. For rotational states built on
vibrations of type $E$ (mixed $S_3$ symmetry) each $L$ state is
twofold degenerate for $K=0$ and fourfold degenerate for $K\neq 0$.
The transformation property of the states under the permutation group
is found by combining the symmetry character of the vibrational and
the rotational wave functions. The rotational  spectrum is obtained by
returning to eq.~(2) and observing that, for a $S_3$-invariant mass operator,
there are four independent terms that determine the rotational spectrum
\cite{BL}. Two of these terms, $ \kappa_{1} \, \hat L \cdot \hat L
+ \kappa_{2} \, \hat K_y^2 $, commute with any $S_3$-invariant operator
and hence correspond to exact symmetries.
Here $\hat L = \sqrt{2} \, (p_{\rho}^{\dagger} \tilde{p}_{\rho}
+p_{\lambda}^{\dagger} \tilde{p}_{\lambda})^{(1)}$ is the angular
momentum operator and
$\hat K_y = -i\sqrt{3} \, (p_{\rho}^{\dagger} \tilde{p}_{\lambda}
-p_{\lambda}^{\dagger} \tilde{p}_{\rho})^{(0)}$
corresponds, for large values of $N$, to a rotation
about the threefold symmetry axis (the $y$-axis in our convention).
The corresponding eigenvalues are
$\kappa_{1} \, L(L+1) + \kappa_{2} \, K^2_y$ and the spectrum is
shown schematically in Fig.~2. The other two rotational terms
do not commute with the vibrational part of $\hat M^{2}_{U(7)}$
and thus induce rotation-vibration couplings \cite{BL}.
Although within the $U(7)$ model we can take these terms into account
as well, we do not consider them here.

The above indicated rotational terms do not lead to a feature of hadronic
spectra expected on the basis of QCD and extensively investigated
decades ago \cite{JT}, namely, the occurrence of linear Regge trajectories.
However, since $L$ and $\vert K_y\vert $ are good quantum numbers, we can
consider, still remaining
with $U(7)$, more complicated functional forms, $f(\hat L^{2}) +
g(\hat K^{2}_{y})$, with eigenvalues $f[L(L+1)] + g[K^{2}_{y}]$. Linear
Regge trajectories can be simply obtained by choosing the form
\ba
\hat M^{2}_{\mbox{rot}}
= \alpha \, \sqrt{\hat L \cdot \hat L + \frac{1}{4}} ~,
\ea
with eigenvalues
\ba
M^{2}_{\mbox{rot}} = \alpha \, (L+1/2) ~.
\ea
where $\alpha$ characterizes the slope of the trajectories.

In comparing with experimental data, the final step is to add the
spin-flavor part. For this part, we take the standard
$SU(6) \supset  SU(3) \otimes SU(2)$ spin-flavor dynamic symmetry of
G\"ursey and Radicati \cite{GR}. In general, this symmetry may be broken
(as it is the case with the hyperfine interaction in the quark potential
model). Here we limit ourselves to the diagonal part
\ba
M^2_{\mbox{spin-flavor}}
= a \, \Bigl [ \langle \hat C_{SU(6)} \rangle - 45 \Bigr ]
+ b \, \Bigl [ \langle \hat C_{SU(3)} \rangle -  9 \Bigr ]
+ c \, \langle \hat C_{SU(2)} \rangle \, ~.
\ea
The first term involves the Casimir operator of the $SU(6)$
spin-flavor group with eigenvalues 45, 33 and 21 for the
representations $56 \leftrightarrow A_1$, $70 \leftrightarrow E$ and
$20 \leftrightarrow A_2$, respectively. The second term
involves the Casimir invariant of the $SU(3)$ flavor group with
eigenvalues 9 and 18 for the octet and decuplet, respectively.
The last term contains the eigenvalues $S(S+1)$ of the spin operator.

Combination of all pieces gives us a closed mass formula for the nonstrange
baryon resonances,
\ba
M^2 = M_{0}^{2} + M^{2}_{\mbox{vib}}
+ M^{2}_{\mbox{rot}} + M^{2}_{\mbox{spin-flavor}} ~.
\ea
This formula contains 7 parameters, $M^2_0$, $\lambda_1 N$, $\lambda_2 N$,
$\alpha$, $a$, $b$ and $c$. We have used it to describe the spectrum of
the $N$ and $\Delta$ families of states, given in Table~1.
The parameters $M^2_0$, $b$, $\lambda_1 N$ and $\lambda_2 N$ have been
determined from the masses of $N(939)$, $\Delta(1232)$, $N(1440)$
and $N(1710)$, respectively. The other parameters $a$, $c$ and
$\alpha$ have been determined by fitting the masses of the remaining
resonances shown in Table~1.
The values of the parameters (in GeV$^2$) are
\ba
& M_{0}^{2}=0.260 ~, \;\; \lambda_1 N=1.192 ~, \;\;
\lambda_2 N=1.538 ~, \;\; \alpha=1.056 ~,
\nonumber\\
& a=-0.042 ~, \;\; b=0.029 ~, \;\; c=0.125 ~.
\ea
The value of the inverse slope of the Regge trajectories ($\alpha = 1.056$
GeV$^{2}$) is consistent with QCD estimates \cite{JT} and identical with that
extracted from an analysis of meson masses \cite{IMZ} ($\alpha = 1.092$
GeV$^{2}$). As one can see from Table~1, the overall fit is of comparable
quality to that of quark potential models \cite{IK,CI}.
We have associated $N(1440)$, $\Delta(1600)$ and
$\Delta(1900)$ with an $A_1$-vibration ($n_u=1, n_v+n_w=0$), and
$N(1710)$ with an $E$-vibration ($n_u=0, n_v+n_w=1$). The remaining states
are rotational excitations belonging to the ground band ($n_u=n_v+n_w=0$).
It is important to note that in the oblate top classification the $N(1440)$
Roper resonance is a one-phonon excitation, whereas in a harmonic
oscillator quark model it is a two-phonon excitation.

In order to emphasize the difference between the model presented here and
the quark potential model, it is worthwhile to compare the respective wave
functions. With the mass operator of eq.~(5) the oblate top wave
functions depend only on $\hat M^{2}_{\mbox{vib}}$ (since the other terms are
diagonal). In particular, the wave functions of the rotational states
belonging to the ground band ($n_u=n_v+n_w=0$) are obtained by
projection from the equilibrium condensate, and therefore depend only
on the value of $R^2$.
In Figure~3, we show the expansion of the ground state wave function
with $L^{\pi}_t = 0^+_S$ (the nucleon $N(939)$) in the harmonic oscillator
basis for $R^2=0.2,\, 1.0$ and $5.0$. The oblate top
wave functions are spread over many oscillator shells and hence are truly
collective in nature. In the harmonic oscillator limit \cite{IK} the wave
function is pure $n=0$. When perturbations are added, the nucleon wave
function acquires small admixtures ($19 \%$ of $n=2$ in \cite{IKK}).

Despite the large differences in wave functions, both the collective oblate top
and the quark potential model give equally good descriptions of the baryon
masses. This indicates that masses alone are not sufficient to distinguish
between these two different scenarios.
Transition form factors are far more sensitive to details in the wave
functions. Preliminary study of helicity amplitudes for
photocouplings ($N^{*}\rightarrow N+\gamma$) and the corresponding
transition form factors indeed show results to this effect.
These results will be presented in a subsequent publication.
We also note that strange baryons can be treated with
$U(7)$ in a similar fashion, except that the $S_3$ symmetry is lowered
to $S_2$ when the mass of one of the constituent parts is
different from the other two. Hyperfine, spin-orbit and other
types of interactions can be treated as well.

In conclusion, we have introduced a unified framework for the description of
the structure of baryons, which is obtained by a bosonic quantization
of the Jacobi coordinates, $\vec{\rho}$ and $\vec{\lambda}$, and their
associated momenta. This method extends previous algebraic treatments
\cite{alg} and allows a straightforward
construction of operators and wave functions with appropriate threefold
permutation symmetry. The framework provides a tractable computational
scheme in which all calculations can be done exactly. It encompasses both
single-particle and collective forms of dynamics so that
different models for the structure of baryons can be studied and compared.
We have analyzed a particular collective model in which baryon
resonances are treated as rotational and vibrational excitations of an oblate
symmetric top. This model provides a description of baryon masses
of comparable quality to that of quark potential models. Our results
indicate that baryon masses are mostly determined by simple features of
strong interactions, namely, the threefold permutation symmetry
and the flavor-spin symmetry.

This work is supported in part by the Stichting voor Fundamenteel Onderzoek
der Materie (FOM) with financial support from the Nederlandse Organisatie
voor Wetenschappelijk Onderzoek (NWO), the U.S. Department of Energy Grant
DE-FG-02-91ER40608 and the Basic Research Foundation of the
Israel Academy of Sciences and Humanities.

\clearpage

\begin{figure}
\centering
\setlength{\unitlength}{1pt}
\begin{picture}(320,240)(-0,-30)
\end{picture}
\vspace{5truecm}
\caption[]{\small Geometric arrangement for baryons. \normalsize}
\end{figure}

\clearpage

\begin{figure}
\setlength{\unitlength}{1pt}
\begin{picture}(400,210)(-25,-10)
\thinlines
\put (  0,-10) {\line(1,0){400}}
\put (  0,200) {\line(1,0){400}}
\put (  0,-10) {\line(0,1){210}}
\put (185,-10) {\line(0,1){210}}
\put (400,-10) {\line(0,1){210}}
\thicklines
\put (10, 50) {\line(1,0){20}}
\put (10, 70) {\line(1,0){20}}
\put (10,110) {\line(1,0){20}}
\put (10,170) {\line(1,0){20}}
\put (50, 66) {\line(1,0){20}}
\put (50,106) {\line(1,0){20}}
\put (50,166) {\line(1,0){20}}
\put (90, 94) {\line(1,0){20}}
\put (90,154) {\line(1,0){20}}
\put (130,134) {\line(1,0){20}}
\thinlines
\put ( 53,  5) {$A_1$-type vibration}
\put ( 10, 25) {$K$=0}
\put ( 50, 25) {$K$=1}
\put ( 90, 25) {$K$=2}
\put (130, 25) {$K$=3}
\put ( 32, 50) {$0^+_{A_1}$}
\put ( 32, 70) {$1^+_{A_2}$}
\put ( 32,110) {$2^+_{A_1}$}
\put ( 32,170) {$3^+_{A_2}$}
\put ( 72, 66) {$1^-_{E}$}
\put ( 72,106) {$2^-_{E}$}
\put ( 72,166) {$3^-_{E}$}
\put (112, 94) {$2^+_{E}$}
\put (112,154) {$3^+_{E}$}
\put (152,134) {$3^-_{A_1 A_2}$}
\thicklines
\put (195, 50) {\line(1,0){20}}
\put (195, 70) {\line(1,0){20}}
\put (195,110) {\line(1,0){20}}
\put (195,170) {\line(1,0){20}}
\put (235, 66) {\line(1,0){20}}
\put (235,106) {\line(1,0){20}}
\put (235,166) {\line(1,0){20}}
\put (295, 94) {\line(1,0){20}}
\put (295,154) {\line(1,0){20}}
\put (355,134) {\line(1,0){20}}
\thinlines
\put (253,  5) {$E$-type vibration}
\put (195, 25) {$K$=0}
\put (235, 25) {$K$=1}
\put (295, 25) {$K$=2}
\put (355, 25) {$K$=3}
\put (217, 50) {$0^+_E$}
\put (217, 70) {$1^+_E$}
\put (217,110) {$2^+_E$}
\put (217,170) {$3^+_E$}
\put (257, 66) {$1^-_{A_1 A_2 E}$}
\put (257,106) {$2^-_{A_1 A_2 E}$}
\put (257,166) {$3^-_{A_1 A_2 E}$}
\put (317, 94) {$2^+_{A_1 A_2 E}$}
\put (317,154) {$3^+_{A_1 A_2 E}$}
\put (377,134) {$3^-_{E E}$}
\end{picture}
\caption[]{\small
Schematic representation of the rotational structure built on
vibrations of type $A_1$ (lefthand side) and vibrations of type $E$
(righthand side). The levels are labeled by $K$, $L^{\pi}_t$, where
$t$ denotes the overall (vibrational plus rotational) permutation
symmetry. Each $E$ state is doubly degenerate.
\normalsize}
\end{figure}
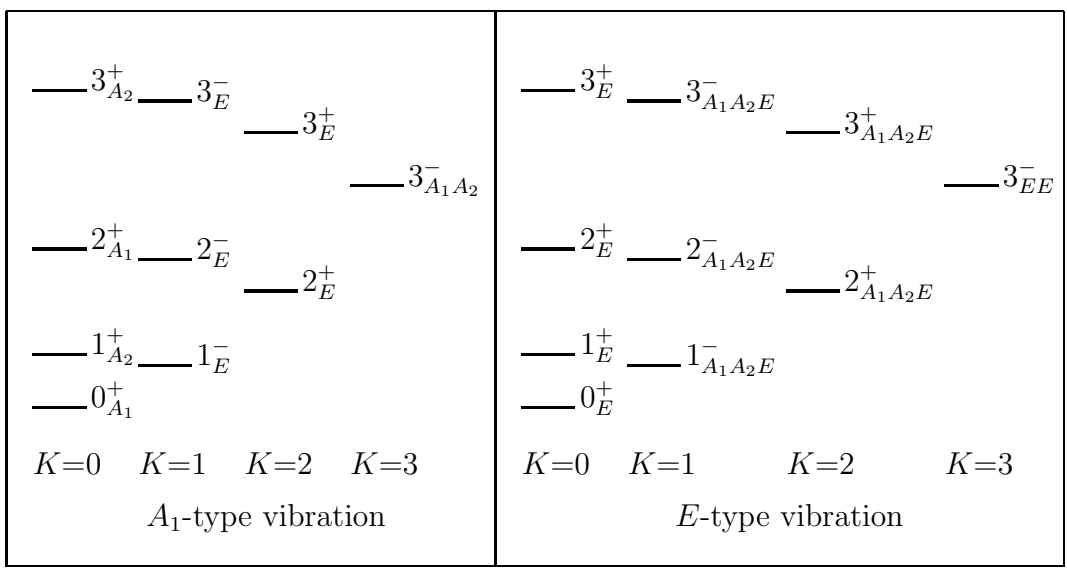

\clearpage

\begin{figure}
\centering
\setlength{\unitlength}{1pt}
\begin{picture}(320,240)(-0,-30)
\thinlines
\put (  0,  0) {\line(1,0){320}}
\put (  0,  0) {\line(0,1){240}}
\put (  0,240) {\line(1,0){320}}
\put (320,  0) {\line(0,1){240}}
\put ( 10,  0) {\line(0,1){5}}
\put (110,  0) {\line(0,1){5}}
\put (210,  0) {\line(0,1){5}}
\put (310,  0) {\line(0,1){5}}
\put ( 10,-20) {0}
\put (105,-20) {10}
\put (160,-30) {\large $n$ \normalsize}
\put (205,-20) {20}
\put (305,-20) {30}
\put (  0, 20) {\line(1,0){5}}
\put (  0,120) {\line(1,0){5}}
\put (  0,220) {\line(1,0){5}}
\put (-25, 15) {0.}
\put (-25,115) {0.25}
\put (-40,175) {\large Prob. \normalsize}
\put (-25,215) {0.50}
\put (  0, 20) {\line(1,0){80}}
\put ( 80, 28) {\line(1,0){20}}
\put ( 80, 20) {\line(0,1){ 8}}
\put (100, 56) {\line(1,0){20}}
\put (100, 28) {\line(0,1){28}}
\put (120,104) {\line(1,0){20}}
\put (120, 56) {\line(0,1){48}}
\put (140,132) {\line(1,0){20}}
\put (140,104) {\line(0,1){28}}
\put (160,112) {\line(1,0){20}}
\put (160,112) {\line(0,1){20}}
\put (180, 68) {\line(1,0){20}}
\put (180, 68) {\line(0,1){44}}
\put (200, 32) {\line(1,0){20}}
\put (200, 32) {\line(0,1){36}}
\put (220, 20) {\line(1,0){100}}
\put (220, 20) {\line(0,1){12}}
\put (140,152) {$R^2=1.0$}
\put (  0, 64) {\line(1,0){20}}
\put ( 20,144) {\line(1,0){20}}
\put ( 20, 64) {\line(0,1){80}}
\put ( 40,160) {\line(1,0){20}}
\put ( 40,144) {\line(0,1){16}}
\put ( 60, 92) {\line(1,0){20}}
\put ( 60, 92) {\line(0,1){68}}
\put ( 80, 40) {\line(1,0){20}}
\put ( 80, 40) {\line(0,1){52}}
\put (100, 24) {\line(1,0){20}}
\put (100, 24) {\line(0,1){16}}
\put (120, 20) {\line(1,0){180}}
\put (120, 20) {\line(0,1){ 4}}
\put ( 40,180) {$R^2=0.2$}
\put (  0, 20) {\line(1,0){200}}
\put (200, 32) {\line(1,0){20}}
\put (200, 20) {\line(0,1){12}}
\put (220, 80) {\line(1,0){20}}
\put (220, 32) {\line(0,1){48}}
\put (240,152) {\line(1,0){20}}
\put (240, 80) {\line(0,1){72}}
\put (260,156) {\line(1,0){20}}
\put (260,152) {\line(0,1){ 4}}
\put (280, 68) {\line(1,0){20}}
\put (280, 68) {\line(0,1){88}}
\put (300, 24) {\line(1,0){20}}
\put (300, 24) {\line(0,1){44}}
\put (240,176) {$R^2=5.0$}
\end{picture}
\caption[]{\small Probability distribution of the ground state
wave function with $L^{\pi}_t = 0^+_S$ in a harmonic oscillator
basis, calculated for $R^2=0.2$, 1.0 and 5.0.
$n=n_{\rho}+n_{\lambda}$ is the total number of oscillator quanta.
The total number of bosons is $N=30$~. \normalsize}
\end{figure}
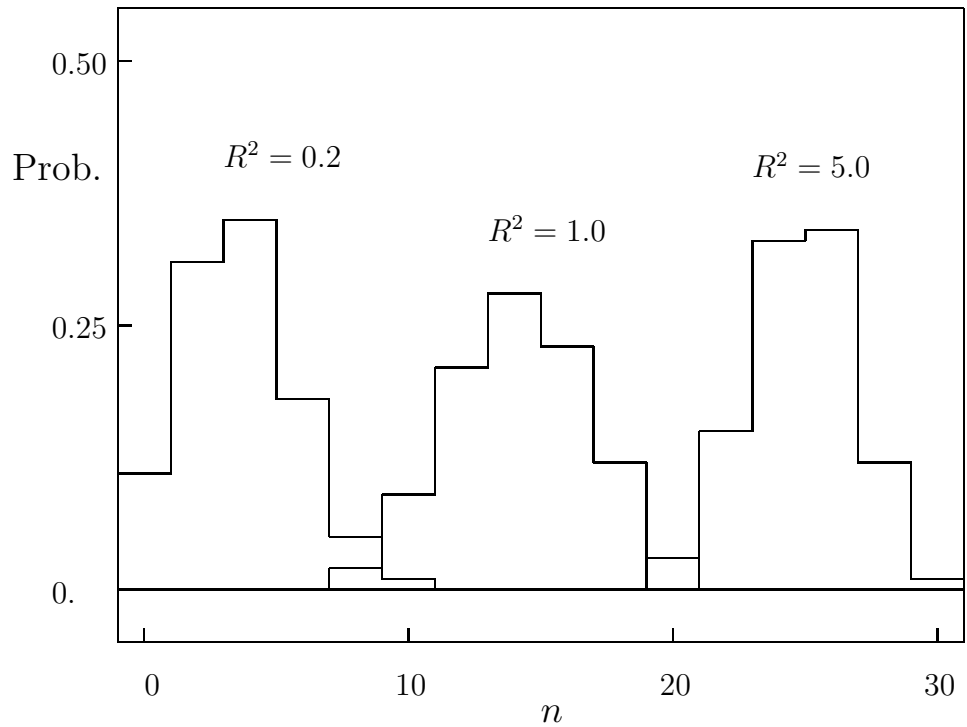

\clearpage

\begin{table}
\centering
\begin{tabular}{llcrccccc}
\hline
& & & & & & & & \\
Baryon & Status & Mass & $J^{\pi}$ & $(n_{u},n_{v}+n_{w})$
& $L^{\pi},K_{y}$ & $S$ & $t$ & $M_{\mbox{calc}}$ \\
& & & & & & & & \\
\hline
& & & & & & & & \\
$N(939)P_{11}$ & **** & 939 & ${1\over 2}^{+}$ & (0,0) &
$0^{+},0$ & ${1\over 2}$ & $A_{1}$ & 939 \\
$N(1440)P_{11}$ & **** & 1430-1470 & ${1\over 2}^{+}$ & (1,0) &
$0^{+},0$ & ${1\over 2}$ & $A_{1}$ & 1440 \\
$N(1520)D_{13}$ & **** & 1515-1530 & ${3\over 2}^{-}$ & (0,0) &
$1^{-},1$ & ${1\over 2}$ & $E$ & 1563 \\
$N(1535)S_{11}$ & **** & 1520-1555 & ${1\over 2}^{-}$ & (0,0) &
$1^{-},1$ & ${1\over 2}$ & $E$ & 1563 \\
$N(1650)S_{11}$ & **** & 1640-1680 & ${1\over 2}^{-}$ & (0,0) &
$1^{-},1$ & ${3\over 2}$ & $E$ & 1678 \\
$N(1675)D_{15}$ & **** & 1670-1685 & ${5\over 2}^{-}$ & (0,0) &
$1^{-},1$ & ${3\over 2}$ & $E$ & 1678 \\
$N(1680)F_{15}$ & **** & 1675-1690 & ${5\over 2}^{+}$ & (0,0) &
$2^{+},0$ & ${1\over 2}$ & $A_{1}$ & 1730 \\
$N(1700)D_{13}$ & *** & 1650-1750 & ${3\over 2}^{-}$ & (0,0) &
$1^{-},1$ & ${3\over 2}$ & $E$ & 1678 \\
$N(1710)P_{11}$ & *** & 1680-1740 & ${1\over 2}^{+}$ & (0,1) &
$0^{+},0$ & ${1\over 2}$ & $E$ & 1710 \\
$N(1720)P_{13}$ & **** & 1650-1750 & ${3\over 2}^{+}$ & (0,0) &
$2^{+},0$ & ${1\over 2}$ & $A_{1}$ & 1730 \\
$N(2190)G_{17}$ & **** & 2100-2200 & ${7\over 2}^{-}$ & (0,0) &
$3^{-},1$ & ${1\over 2}$ & $E$ & 2134 \\
$N(2220)H_{19}$ & **** & 2180-2310 & ${9\over 2}^{+}$ & (0,0) &
$4^{+},0$ & ${1\over 2}$ & $A_{1}$ & 2260 \\
$N(2250)G_{19}$ & **** & 2170-2310 & ${9\over 2}^{-}$ & (0,0) &
$3^{-},1$ & ${3\over 2}$ & $E$ & 2220 \\
$N(2600)I_{1,11}$ & *** & 2550-2750 & ${11\over 2}^{-}$ & (0,0) &
$5^{-},1$ & ${1\over 2}$ & $E$ & 2582 \\
& & & & & & & & \\
\hline
& & & & & & & & \\
$\Delta(1232)P_{33}$ & **** & 1230-1234 & ${3\over 2}^{+}$ & (0,0) &
$0^{+},0$ & ${3\over 2}$ & $A_{1}$ & 1232 \\
$\Delta(1600)P_{33}$ & *** & 1550-1700 & ${3\over 2}^{+}$ & (1,0) &
$0^{+},0$ & ${3\over 2}$ & $A_{1}$ & 1646 \\
$\Delta(1620)S_{31}$ & **** & 1615-1675 & ${1\over 2}^{-}$ & (0,0) &
$1^{-},1$ & ${1\over 2}$ & $E$ & 1644 \\
$\Delta(1700)D_{33}$ & **** & 1670-1770 & ${3\over 2}^{-}$ & (0,0) &
$1^{-},1$ & ${1\over 2}$ & $E$ & 1644 \\
$\Delta(1900)S_{31}$ & *** & 1850-1950 & ${1\over 2}^{-}$ & (1,0) &
$1^{-},1$ & ${1\over 2}$ & $E$ & 1974 \\
$\Delta(1905)F_{35}$ & **** & 1870-1920 & ${5\over 2}^{+}$ & (0,0) &
$2^{+},0$ & ${3\over 2}$ & $A_{1}$ & 1905 \\
$\Delta(1910)P_{31}$ & **** & 1870-1920 & ${1\over 2}^{+}$ & (0,0) &
$2^{+},0$ & ${3\over 2}$ & $A_{1}$ & 1905 \\
$\Delta(1920)P_{33}$ & *** & 1900-1970 & ${3\over 2}^{+}$ & (0,0) &
$2^{+},0$ & ${3\over 2}$ & $A_{1}$ & 1905 \\
$\Delta(1930)D_{35}$ & *** & 1920-1970 & ${5\over 2}^{-}$ & (0,0) &
$2^{-},1$ & ${1\over 2}$ & $E$ & 1939 \\
$\Delta(1950)F_{37}$ & **** & 1940-1960 & ${7\over 2}^{+}$ & (0,0) &
$2^{+},0$ & ${3\over 2}$ & $A_{1}$ & 1905 \\
$\Delta(2420)H_{3,11}$ & **** & 2300-2500 & ${11\over 2}^{+}$ & (0,0) &
$4^{+},0$ & ${3\over 2}$ & $A_{1}$ & 2396 \\
& & & & & & & & \\
\hline
& & & & & & & &
\end{tabular}
\caption[]{\small Oblate top classification
of the nucleon and the delta families of baryon resonances.
$t$ denotes the overall permutation symmetry. The masses are given in MeV.
The experimental values are taken from \cite{PDG}.
The average r.m.s. deviation is 38 MeV.
\normalsize}
\end{table}

\end{document}